\PassOptionsToPackage{unicode}{hyperref}
\PassOptionsToPackage{hyphens}{url}
\PassOptionsToPackage{dvipsnames,svgnames,x11names}{xcolor}
\documentclass[
  letterpaper,
  DIV=11,
  numbers=noendperiod]{scrartcl}
\usepackage{xcolor}
\usepackage{amsmath,amssymb}
\usepackage{iftex}
\ifPDFTeX
  \usepackage[T1]{fontenc}
  \usepackage[utf8]{inputenc}
  \usepackage{textcomp} % provide euro and other symbols
\else % if luatex or xetex
  \usepackage{unicode-math} % this also loads fontspec
  \defaultfontfeatures{Scale=MatchLowercase}
  \defaultfontfeatures[\rmfamily]{Ligatures=TeX,Scale=1}
\fi
\usepackage{lmodern}
\ifPDFTeX\else
\fi
\IfFileExists{upquote.sty}{\usepackage{upquote}}{}
\IfFileExists{microtype.sty}{% use microtype if available
  \usepackage[]{microtype}
  \UseMicrotypeSet[protrusion]{basicmath} % disable protrusion for tt fonts
}{}
\makeatletter
\@ifundefined{KOMAClassName}{% if non-KOMA class
  \IfFileExists{parskip.sty}{%
    \usepackage{parskip}
  }{% else
    \setlength{\parindent}{0pt}
    \setlength{\parskip}{6pt plus 2pt minus 1pt}}
}{% if KOMA class
  \KOMAoptions{parskip=half}}
\makeatother
\makeatletter
\ifx\paragraph\undefined\else
  \let\oldparagraph\paragraph
  \renewcommand{\paragraph}{
    \@ifstar
      \xxxParagraphStar
      \xxxParagraphNoStar
  }
  \newcommand{\xxxParagraphStar}[1]{\oldparagraph*{#1}\mbox{}}
  \newcommand{\xxxParagraphNoStar}[1]{\oldparagraph{#1}\mbox{}}
\fi
\ifx\subparagraph\undefined\else
  \let\oldsubparagraph\subparagraph
  \renewcommand{\subparagraph}{
    \@ifstar
      \xxxSubParagraphStar
      \xxxSubParagraphNoStar
  }
  \newcommand{\xxxSubParagraphStar}[1]{\oldsubparagraph*{#1}\mbox{}}
  \newcommand{\xxxSubParagraphNoStar}[1]{\oldsubparagraph{#1}\mbox{}}
\fi
\makeatother

\usepackage{longtable,booktabs,array}
\usepackage{calc} % for calculating minipage widths
\usepackage{etoolbox}
\makeatletter
\patchcmd\longtable{\par}{\if@noskipsec\mbox{}\fi\par}{}{}
\makeatother
\IfFileExists{footnotehyper.sty}{\usepackage{footnotehyper}}{\usepackage{footnote}}
\makesavenoteenv{longtable}
\usepackage{graphicx}
\makeatletter
\newsavebox\pandoc@box
\newcommand*\pandocbounded[1]{% scales image to fit in text height/width
  \sbox\pandoc@box{#1}%
  \Gscale@div\@tempa{\textheight}{\dimexpr\ht\pandoc@box+\dp\pandoc@box\relax}%
  \Gscale@div\@tempb{\linewidth}{\wd\pandoc@box}%
  \ifdim\@tempb\p@<\@tempa\p@\let\@tempa\@tempb\fi% select the smaller of both
  \ifdim\@tempa\p@<\p@\scalebox{\@tempa}{\usebox\pandoc@box}%
  \else\usebox{\pandoc@box}%
  \fi%
}
\def\fps@figure{htbp}
\makeatother

\NewDocumentCommand\citeproctext{}{}

\makeatletter
 \let\@cite@ofmt\@firstofone
 \def\@biblabel#1{}
 \def\@cite#1#2{{#1\if@tempswa , #2\fi}}
\makeatother
\newlength{\cslhangindent}
\newlength{\csllabelwidth}
\newenvironment{CSLReferences}[2] % #1 hanging-indent, #2 entry-spacing
 {\begin{list}{}{%
  \setlength{\itemindent}{0pt}
  \setlength{\leftmargin}{0pt}
  \setlength{\parsep}{0pt}
  \ifodd #1
   \setlength{\leftmargin}{\cslhangindent}
   \setlength{\itemindent}{-1\cslhangindent}
  \fi
  \setlength{\itemsep}{#2\baselineskip}}}
 {\end{list}}
\usepackage{calc}

\usepackage{booktabs}
\usepackage{longtable}
\usepackage{array}
\usepackage{multirow}
\usepackage{wrapfig}
\usepackage{float}
\usepackage{colortbl}
\usepackage{pdflscape}
\usepackage{tabu}
\usepackage{threeparttable}
\usepackage{threeparttablex}
\usepackage[normalem]{ulem}
\usepackage{makecell}
\usepackage{xcolor}
\usepackage{float}
\KOMAoption{captions}{tableheading}
\makeatletter
\@ifpackageloaded{caption}{}{\usepackage{caption}}
\AtBeginDocument{%
\ifdefined\contentsname
  \renewcommand*\contentsname{Table of contents}
\else
  \newcommand\contentsname{Table of contents}
\fi
\ifdefined\listfigurename
  \renewcommand*\listfigurename{List of Figures}
\else
  \newcommand\listfigurename{List of Figures}
\fi
\ifdefined\listtablename
  \renewcommand*\listtablename{List of Tables}
\else
  \newcommand\listtablename{List of Tables}
\fi
\ifdefined\figurename
  \renewcommand*\figurename{Figure}
\else
  \newcommand\figurename{Figure}
\fi
\ifdefined\tablename
  \renewcommand*\tablename{Table}
\else
  \newcommand\tablename{Table}
\fi
}
\@ifpackageloaded{float}{}{\usepackage{float}}
\floatstyle{ruled}
\@ifundefined{c@chapter}{\newfloat{codelisting}{h}{lop}}{\newfloat{codelisting}{h}{lop}[chapter]}
\floatname{codelisting}{Listing}

\makeatother
\makeatletter
\@ifpackageloaded{caption}{}{\usepackage{caption}}
\@ifpackageloaded{subcaption}{}{\usepackage{subcaption}}
\makeatother
\usepackage{bookmark}
\IfFileExists{xurl.sty}{\usepackage{xurl}}{} % add URL line breaks if available
\hypersetup{
  pdftitle={Benchmarking AI Performance on End-to-End Data Science Projects},
  pdfauthor={Evelyn Hughes; Rohan Alexander},
  colorlinks=true,
  linkcolor={blue},
  filecolor={Maroon},
  citecolor={Blue},
  urlcolor={Blue},
  pdfcreator={LaTeX via pandoc}}

\title{Benchmarking AI Performance on End-to-End Data Science
Projects\thanks{We gratefully acknowledge the financial support of the
Data Sciences Institute at the University of Toronto, NSERC Alliance
(110\_2024\_2025\_Q4\_13), and SSHRC Partnership. We thank Monica
Alexander, Alice Malmberg, and participants at the 2026 Southern
Political Science Association Conference for helpful suggestions. Hughes
conducted this analysis while an intern at the Investigative Journalism
Foundation. Contact:
\href{mailto:rohan.alexander@utoronto.ca}{\nolinkurl{rohan.alexander@utoronto.ca}}.}}
\author{Evelyn Hughes \and Rohan Alexander}
\date{February 15, 2026}
\begin{document}
\maketitle
\begin{abstract}
Data science is an integrated workflow of technical, analytical,
communication, and ethical skills, but current AI benchmarks focus
mostly on constituent parts. We test whether AI models can generate
end-to-end data science projects. To do this we create a benchmark of 40
end-to-end data science projects with associated rubric evaluations. We
use these to build an automated grading pipeline that systematically
evaluates the data science projects produced by generative AI models. We
find the extent to which generative AI models can complete end-to-end
data science projects varies considerably by model. Most recent models
did well on structured tasks, but there were considerable differences on
tasks that needed judgment. These findings suggest that while AI models
could approximate entry-level data scientists on routine tasks, they
require verification.
\end{abstract}

\newpage


\textbf{Corresponding author:} Rohan Alexander:
\href{mailto:rohan.alexander@utoronto.ca}{\nolinkurl{rohan.alexander@utoronto.ca}}.

\textbf{CRediT contributions:} \emph{Hughes:} Data Curation; Formal
Analysis; Investigation; Methodology; Software; Visualization; Writing
-- Original Draft Preparation; Writing -- Review \& Editing.
\emph{Alexander:} Conceptualization; Funding Acquisition; Investigation;
Methodology; Project Administration; Resources; Supervision;
Visualization; Writing -- Original Draft Preparation; Writing -- Review
\& Editing.

\textbf{AI usage:} We used AI to help with aspects of coding including
interacting with APIs, data cleaning, and creating figures and tables,
but the authors reviewed and evaluated all code and results. We used AI
to suggest related literature and review the drafted paper to identify
mistakes, inconsistencies, and awkward phrasing.

\textbf{Significance statement:} Existing benchmarks evaluate AI models
on isolated coding or statistical tasks, but real-world data science
requires integrated workflows that bring together tasks such as question
formulation, data collection, analysis, visualization, and
communication. We develop a benchmark assessing whether AI models can
complete end-to-end data science projects that meet undergraduate
coursework standards. We find differences between models: the recent
model Claude Opus 4.6 scored 85 per cent while the older model GPT-4o
scored 32 per cent, compared with an undergraduate A-/B+ threshold of 80
per cent. Models were better at template-following tasks than with tasks
that required judgment such as evaluating measurement, correct citation
practices, and writing abstracts. AI models could approximate
entry-level data scientists on routine tasks but require verification.

\textbf{Teaser:} AI can match good undergraduates on end-to-end data
science projects but varies more on tasks needing judgment.

\newpage

\section{Introduction}\label{introduction}

Data science is ``the process of generating insight from data through
reproducible and auditable processes'' (Timbers, Campbell, and Lee 2022,
xvii). There is much more to data science than just writing code to
build models. It requires blending a variety of skills including:
mathematical, computational, and statistical foundations; data
management, curation, visualization, modeling and assessment; workflow
and reproducibility; communication and teamwork; domain-specific
knowledge; and ethics (National Academies of Sciences, Engineering, and
Medicine 2018, 22). Each skill requires making technical and
non-technical decisions, any of which could affect the overall outcome.
Data science requires putting these skills together in a workflow that
shows high-quality judgment. This broader focus makes a benchmark
focused not just on constituent skills, but also on the broader
reproducible workflow, important.

There is considerable interest in the effect of generative AI on
knowledge work, of which data science is an instance. AI has been found
to intensify work by expanding task scope and accelerating pace
(Ranganathan and Ye 2026) and to increase productivity in an
experimental setting (Noy and Zhang 2023). AI is affecting both the
quality and quantity of scientific work (Kusumegi et al. 2025; Kobak et
al. 2025). \texttt{DataSciBench} is a benchmark of data science tasks
that includes cleaning, exploration, visualization, and predictive
modeling and uses predefined prompts with automated metrics to evaluate
performance (Zhang et al. 2025). But this may not capture the judgment
required in real-world analysis where data scientists must formulate
questions and make decisions without explicit instructions. Similarly,
focusing on code generation for data (Lai et al. 2023; Zhuo et al.
2025); evaluating AI performance in machine learning and econometrics
(Gu et al. 2025; Q. Chen et al. 2025); or looking at specific data
analysis tasks (Evkaya and Carvalho 2026), does not necessarily consider
the workflow and communication aspects that are critical for data
science and many other knowledge work jobs.

Understanding where AI succeeds and fails on integrated workflows is
important to understand its potential economic impacts. We evaluate
generative AI on end-to-end data science projects. Our benchmark tests
the extent to which generative AI models can develop questions, code in
Python, clean and prepare data, create visualizations, and write a
paper. This is closely related to the work of professional data
scientists (Robinson and Nolis 2020, 5).

We construct a benchmark of 40 projects that university undergraduates
initially wrote. Two human experts evaluated each project on a
17-category rubric totaling 45 points. The A-/B+ threshold for
undergraduates on these projects is 80 per cent. We use that benchmark
to develop an automated grading pipeline using LLM-as-a-judge. We then
use that to evaluate five projects from each of seven AI models.

Overall, on average, Anthropic's Claude Opus 4.6 got 85 per cent;
OpenAI's GPT-5.2 got 82 per cent; Google's Gemini 3 Flash got 78 per
cent; xAI's Grok-4 got 70 per cent; Google's Gemini 3 Pro got 59 per
cent; Meta's Llama 4 got 51 per cent; and OpenAI's older GPT-4o model
got 32 per cent. Scores varied between the different rubric categories.
Models tended to do better on structured tasks, such as writing titles
and test suites. There was larger variance on tasks that needed
judgment, such as writing about measurement, writing an abstract, and
including correct citations.

\section{Results}\label{sec-results}

We evaluated seven AI models: Anthropic's Claude Opus 4.6, Google's
Gemini 3 Flash and Gemini 3 Pro, OpenAI's GPT-5.2 and GPT-4o, xAI's
Grok-4, and Meta's Llama 4. Each model generated five projects. The
auto-grader then graded each project three times against a rubric
(available in Appendix~\ref{sec-rubric}). We conducted analysis using
\texttt{R} (R Core Team 2025) and the \texttt{tidyverse} packages
(Wickham et al. 2019).

\subsection{Overall performance}\label{sec-overall}

Claude Opus 4.6 achieved the highest average score of 38 points (85 per
cent), followed by GPT-5.2 at 37 points (82 per cent), Gemini 3 Flash at
35 points (78 per cent), Grok-4 at 31 points (70 per cent), Gemini 3 Pro
at 27 points (59 per cent), Llama 4 at 23 points (51 per cent), and
GPT-4o at 15 points (32 per cent) (Figure~\ref{fig-total-scores}). These
are averaged across each project (individual project scores are in
Appendix~\ref{sec-per-project}). In Figure~\ref{fig-total-scores} the
dashed line at 36 points is the 80 per cent cut-off between an A- and a
B+. Claude Opus 4.6, GPT-5.2, and Gemini 3 Flash are all around this
level of performance. Grok-4, Gemini 3 Pro, Llama 4, and the older
GPT-4o model, are below it.

\begin{figure}

\centering{

\pandocbounded{\includegraphics[keepaspectratio]{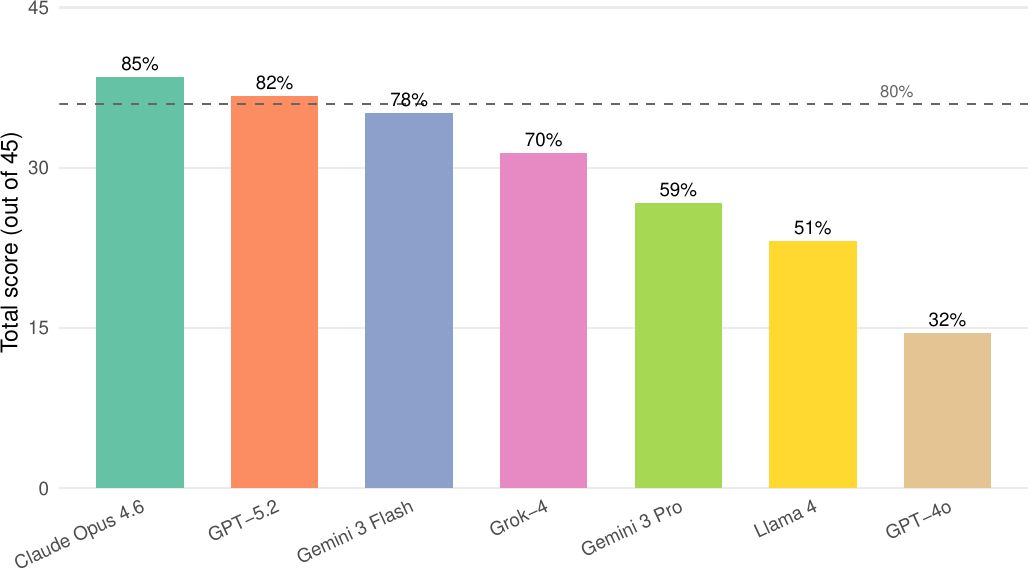}}

}

\caption{\label{fig-total-scores}Overall performance, both score and
percentage, of seven AI models on an end-to-end data analysis project.
The maximum score is 45, and the horizontal dashed line at 36 (80 per
cent) is the threshold between a B+ and A-, which is the typical
performance of good upper-year undergraduates.}

\end{figure}%

\subsection{Performance by category type}\label{sec-category-type}

\begin{figure}

\centering{

\pandocbounded{\includegraphics[keepaspectratio]{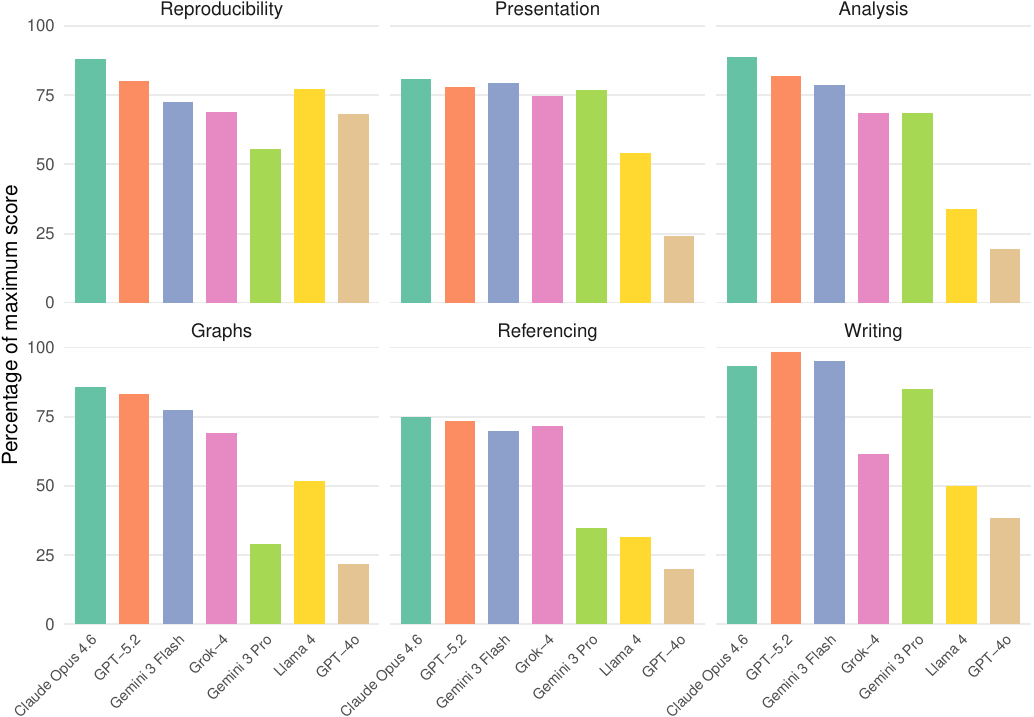}}

}

\caption{\label{fig-type-scores}Model performance by rubric evaluation
type shown as the percentage of maximum possible points for that type.
Evaluation types and maximum number of points are: Reproducibility (9
points), Presentation (10 points), Analysis (10 points), Graphs (8
points), Referencing (4 points), and Writing (4 points).}

\end{figure}

Figure~\ref{fig-type-scores} shows scores aggregated by evaluation type.
Claude Opus 4.6 did well across most types, followed by GPT-5.2. The
older model, GPT-4o, underperformed across all types, scoring zero on
several categories.

Reproducibility (9 points) considers code quality and documentation.
Claude Opus 4.6 achieved 88 per cent, followed by GPT-5.2 at 80 per
cent. Good projects tended to have extensive test suites and detailed
READMEs. Gemini 3 Pro achieved 56 per cent, often due to worse
documentation. Writing (4 points) showed less difference between models:
GPT-5.2 achieved near-full marks including avoiding buzzwords, while
other models lost points for using terms like ``crucial'' and
``insights''. This tests direct instruction-following because the rubric
included the list of words to avoid.

\subsection{Performance by rubric task}\label{sec-category-detail}

\begin{figure}

\centering{

\pandocbounded{\includegraphics[keepaspectratio]{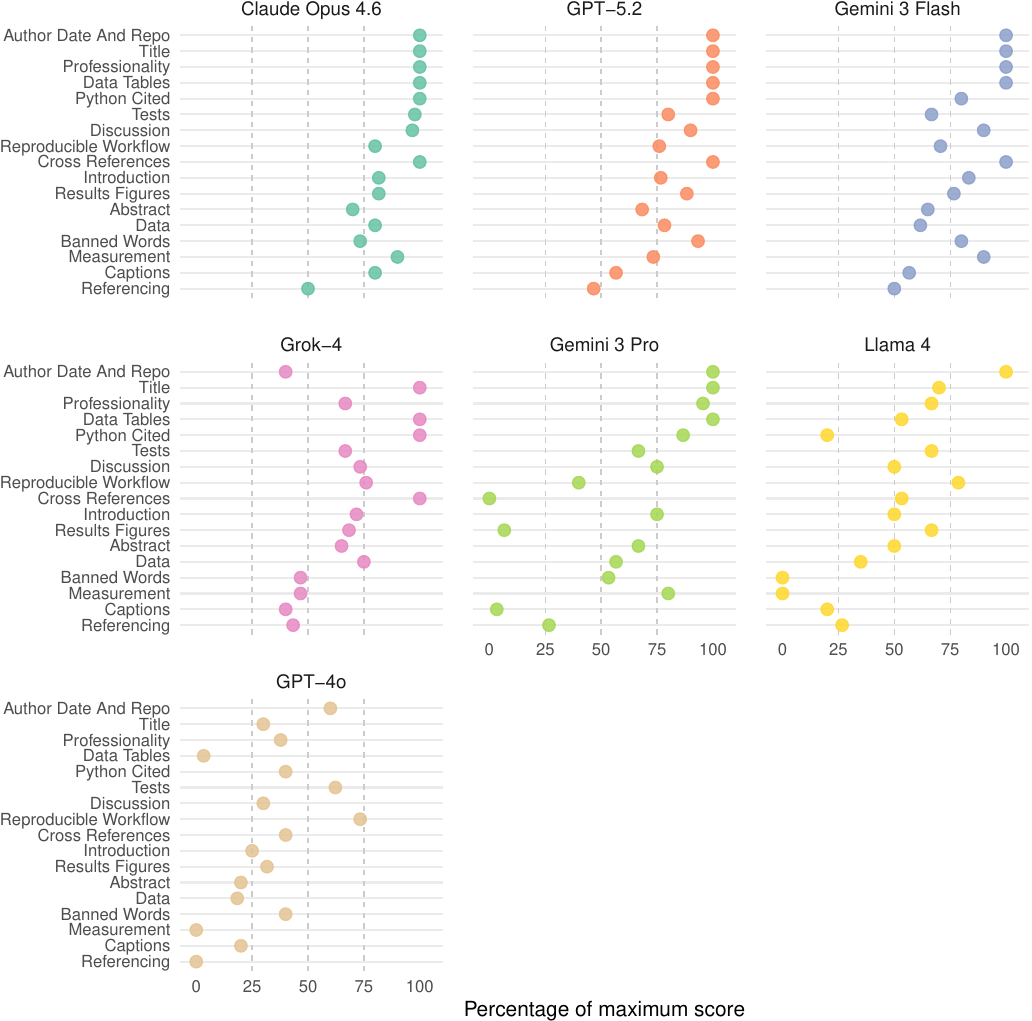}}

}

\caption{\label{fig-category-scores}Model performance for all 17 rubric
categories shown as percentage of maximum possible points for that
category. Categories are ordered by average performance across all
models (highest at top). Each point is one model's average performance.}

\end{figure}%

When looking at the detailed average scores by rubric item there are
clear areas of strength and weakness (Figure~\ref{fig-category-scores}).
Many models achieved perfect or near-perfect scores on Title (2/2). But
more extensive writing requirements, such as the Introduction and
Discussion resulted in differences between models. For instance, Claude
Opus 4.6 scored 3.27/4 on Introduction, while Llama 4 scored 2/4.
Another writing aspect of interest is writing an abstract (4 points).
This is especially interesting because an abstract is so short yet must
communicate so much detail. No model got full marks. Claude Opus 4.6
scored 2.80/4, while Llama 4 scored 2/4. This suggests that even just
benchmarks focused on automated abstract evaluation could be useful.
Writing about data was another area where there were considerable
differences between models. For instance, Claude Opus 4.6 averaged
3.20/4 and GPT-5.2 averaged 3.13/4, while Gemini 3 Pro averaged 2/4.
Lower-scoring projects tended to not discuss the reasons for variable
selection or cleaning decisions.

It was surprising that not every model got full marks for developing a
test suite (3 points). This required consideration of potential failure
modes for the data and writing tests for them. Claude Opus 4.6 achieved
an average of 2.93/3. Strong test suites considered data types, value
ranges, and null handling, while weak ones tended to be overly generic.

The measurement category required explaining how some real-world
phenomenon becomes a dataset entry. The ability to think deeply about
this is important for data scientists and the lower performance of many
models on this may reflect a failure to put a given dataset in context.
It was also surprising that Gemini 3 Pro received zero marks for
cross-references because cross-referencing is well-documented and uses
code.

The models generally did well on template-following tasks such as
writing a title or putting together a test suite. Tasks that involved
synthesis and context, such as writing an introduction or discussion
section, or writing code to make figures, differentiated models. And
data science specific tasks such as measurement, cross-references and
citations received consistently lower marks, with only a couple of
exceptions. A common referencing failure was generating malformed BibTeX
entries for Python packages, which produced placeholder citations such
as ``matplotlib2023?'' with question marks instead of publication
details.

Examples of the first pages of the papers produced by different models
are in Appendix~\ref{sec-examples}.

\section{Discussion}\label{sec-discussion}

There is considerable interest in the extent to which AI will replace
white-collar work. Our findings suggest even the best current AI models
still require human oversight. Claude Opus 4.6 and some other models did
well, but still had notable weaknesses. Surprisingly, Gemini 3 Pro
scored lower than Gemini 3 Flash. Gemini 3 Pro required more retry
attempts during project generation, which may have decreased output
quality. But comparing the two models from OpenAI, one a recent model
(GPT-5.2) and the other an older model (GPT-4o), it is possible that
continued development, or data science specific models, could further
increase performance. It also points to the need for more
domain-specific benchmarks as we examine AI performance on specific
areas of work. But at the moment, an organization interested in
deploying AI for data science cannot simply assume that just any model
could be used for the task without considerable evaluation and
verification.

Models tended to do well when they could follow well-defined templates
or patterns such as Title, and Author/Date/Repo, and do worse in
sections such as Abstract, Introduction, and Discussion. Papers written
by models tended to do worse in these sections not because they were
poorly written---they were not---but because that writing tended to not
communicate much information. Despite the presence of the ``right
words'', they did not seem to communicate deep expertise and thoughtful
reflection. Again, task specific benchmarks may help improve this.

Measurement required describing how some phenomena become an entry in a
dataset. The difficulties that many models had on this may reflect it
being underrepresented in training corpora despite being critical in
data science. Similarly, lower marks on cross-references suggests a
failure to understand Quarto syntax even though the instructions
referenced material with explicit examples (Alexander 2023) and the
provided example. While difficulties with a specialized tool like Quarto
may not be specifically concerning, it may telegraph difficulties when
translating performance on general benchmarks into delivering return on
investment for businesses.

The best models, Claude Opus 4.6, GPT-5.2, and Gemini 3 Flash, scored
around the same as a good undergraduate. Models may soon be able to
complete routine data analysis with minimal revision, but the systematic
weaknesses we document would mean users still need to check them. Fully
autonomous deployment for data science may carry substantial risk,
particularly with lower-performing models.

We did not prompt engineer our models for dataset-specific analysis.
Better performance would be possible with this, but this may undermine
real-world value. Data scientists try to learn from data, conducting
different analyses on different datasets. That said, with further prompt
engineering we may be able to remove some of the guardrails we needed,
such as breaking down the steps into separate API calls and providing
the model with a choice from a handful of datasets. The evaluation of
visualization is difficult and taste is difficult to evaluate. The high
variance we find in figure-related categories may partly reflect
evaluation limitations.

LLM-as-a-judge introduces a variety of biases (Zhu, Wang, and Wang
2025). Of particular concern is whether a grading model favors output
from its own provider. Our primary grader is Anthropic's Claude Sonnet
4.5, which also grades Anthropic's Claude Opus 4.6, the highest-scoring
model. To mitigate this we validated our auto-grader against the
human-graded benchmark using both Claude Sonnet 4.5 and GPT-5.2, and
obtained consistent results (details about grading consistency results
are in Appendix~\ref{sec-grading}). Even if this bias were present and
Claude Opus 4.6 were performing, say, only on par with recent models
from Google and OpenAI, that level of performance would still be
comparable to that of a good undergraduate. While our rubric emphasizes
academic aspects that businesses may be less concerned with, citation
checking is a relatively easy way to detect hallucinations, so we feel
it is still valuable. We also focus on evaluating analysis conducted on
data from the City of Toronto, and findings may be different in other
domains.

Our sample is small and this meant we did not use statistical methods to
look for differences between models. Our finding that, say, Anthropic's
Claude Opus 4.6 was the best and that OpenAI's GPT-5.2 was second should
be considered indicative of the general performance level of recent
models, rather than a definitive ranking.

Future work should continue to increase the number of projects per model
and reduce the guardrails our paper needed. It would also be interesting
to consider how models could perform when used alongside human guidance.

\section{Methods}\label{sec-methods}

We want to systematically evaluate the extent to which AI can complete
data science projects. To do this we first establish a benchmark, then
we use that benchmark to develop and validate an auto-grader. We then
prompt models to generate projects. Finally, we use the auto-grader to
evaluate those in a scalable way.

\subsection{Benchmark construction}\label{benchmark-construction}

Benchmarking is one way to evaluate model capabilities. A benchmark
brings together a dataset of tasks and a performance metric (Reuel et
al. 2024). There are legitimate criticisms of being solely focused on
benchmark performance (Raji et al. 2021; Bowman and Dahl 2021),
especially concern about selective disclosure (Singh et al. 2025).
Nonetheless, evaluation against a benchmark, particularly over time, is
a widely used way to track AI progress.

OpenAI publicly released GPT-3 in 2020, catalyzing considerable focus on
publicly sharing benchmarks around AI code-generation. For instance,
\texttt{HumanEval} (M. Chen et al. 2021) evaluated the capacity of
models to write Python functions from docstrings. Similarly,
\texttt{MBPP} (Austin et al. 2021) was a larger benchmark consisting
mostly of Python programming problems. More recently, \texttt{SWE-bench}
(Jimenez et al. 2024) measures AI progress on writing Python code to
resolve real GitHub issues. These, and similar, benchmarks have various
degrees of complexity and real-world relevance, but all evaluate the
capacity of models to generate code within a particular setting, rather
than measuring capacity to complete a project end-to-end.

To construct our benchmark we collect 40 projects initially completed by
university students and anonymize them. Two human experts then use a
17-category rubric to mark each project and resolve any differences.

The task is to complete a data analysis project using data from Open
Data Toronto. We adapted this task from an upper-year undergraduate
assignment at the University of Toronto. Students can pick any dataset
they want from Open Data Toronto. Students work individually and submit
a link to a repository that includes code and a short paper telling a
story about the data. The entire work needs to be fully reproducible. We
provide students with an example GitHub repo. This also provides more
information about the steps they should take. We also provide detail
about what aspects should go into the paper. In addition to the detailed
instructions, students also receive the detailed rubric we use to mark
their submission. And students can also see past examples.

Each project must include a Python script, \texttt{01-download.py}, that
downloads a dataset from Open Data Toronto. A second script,
\texttt{02-clean.py}, cleans and prepares the data. Students then test
the analysis dataset, \texttt{03-test.py}, to check there are no obvious
issues. They usually do this using the Python \texttt{pydantic} package.
Students write the paper as a Quarto document, \texttt{paper.qmd}, which
compiles to a PDF, \texttt{paper.pdf}. The Quarto document integrates
the code required to generate graphs and tables with text. The required
sections of the paper are: Introduction, Data, Results, and Discussion.
Students must cross-reference all graphs and tables. The paper also
needs to include citations that compile using a BibTeX file,
\texttt{references.bib}. Finally, students must ensure reproducibility
by including a UV lock file and \texttt{pyproject.toml} or
\texttt{requirements.txt}, a detailed \texttt{README.md}, and a
well-structured repository.

We obtained project repositories from 40 students by asking them,
individually, if they were willing to let us use their repository for
this purpose. We did this after the class ended and we finalized and
released grades. Students could complete their projects using either
\texttt{R} or \texttt{Python}. To make our evaluation process easier and
more consistent, we converted projects that originally used \texttt{R}
to \texttt{Python} by making the minimal number of changes needed. This
involved changing the packages and main functions such as from
\texttt{dplyr} to \texttt{pandas}, or from \texttt{ggplot2} to
\texttt{matplotlib}. We anonymized the projects by changing the name and
emails.

For each project two human graders independently evaluated it against a
detailed rubric. The rubric has 17 categories, totaling 45 points, and
we combine them into six groups: Reproducibility, Presentation,
Analysis, Graphs, Referencing, and Writing
(Table~\ref{tbl-rubric-summary}). Each category has points associated
with it and detailed instructions about what is required for each level.
When the markers disagreed they resolved differences through discussion,
and we added additional detail to the rubric instructions. The full
rubric is available in Appendix~\ref{sec-rubric}.

\begin{table}

\caption{\label{tbl-rubric-summary}Grading rubric for end-to-end data
analysis projects. The rubric has 17 categories grouped into six
evaluation types, totaling 45 points: reproducibility (9 points);
presentation (10 points); analysis (10 points); graphs (8 points);
referencing (4 points); writing (4 points).}

\centering{

\centering
\begin{tabular}[t]{llr}
\toprule
Group & Category & Max points\\
\midrule
 & Reproducible workflow & 5\\
\cmidrule{2-3}
 & Tests & 3\\
\cmidrule{2-3}
\multirow[t]{-3}{*}{\raggedright\arraybackslash Reproducibility} & Author, date, and repo & 1\\
\cmidrule{1-3}
 & Title & 2\\
\cmidrule{2-3}
 & Abstract & 4\\
\cmidrule{2-3}
\multirow[t]{-3}{*}{\raggedright\arraybackslash Presentation} & Introduction & 4\\
\cmidrule{1-3}
 & Measurement & 2\\
\cmidrule{2-3}
 & Data & 4\\
\cmidrule{2-3}
\multirow[t]{-3}{*}{\raggedright\arraybackslash Analysis} & Discussion & 4\\
\cmidrule{1-3}
 & Data tables & 2\\
\cmidrule{2-3}
 & Results figures & 4\\
\cmidrule{2-3}
\multirow[t]{-3}{*}{\raggedright\arraybackslash Graphs} & Captions & 2\\
\cmidrule{1-3}
 & Cross-references & 1\\
\cmidrule{2-3}
 & References & 2\\
\cmidrule{2-3}
\multirow[t]{-3}{*}{\raggedright\arraybackslash Referencing} & Python cited & 1\\
\cmidrule{1-3}
 & Professionalism & 3\\
\cmidrule{2-3}
\multirow[t]{-2}{*}{\raggedright\arraybackslash Writing} & Banned words & 1\\
\bottomrule
\end{tabular}

}

\end{table}%

\subsection{Auto-grader development and
validation}\label{auto-grader-development-and-validation}

Assessing the projects created by the generative AI models in a
consistent, scalable way required developing and validating an
auto-grader.

By way of background, the question of how to evaluate responses to the
tasks in a benchmark against the desired response is a common problem
with benchmark evaluation. For open-ended tasks it can be difficult to
uniquely determine what the answer should be and so researchers commonly
define rubrics. There are two ways to evaluate submissions against a
rubric: human-evaluation or LLM-as-a-judge. The latter has the benefit
of being scalable, and with appropriate set-up can approximate human
preferences (Zheng et al. 2023). That said, there are systematic biases
from LLM-as-a-judge, including position and verbosity bias (Zhu, Wang,
and Wang 2025) and particular concerns when specialized knowledge is
required (Szymanski et al. 2024).

We use an automated approach because it allows us to evaluate many more
submissions from the generative AI models than would be viable using
human evaluation. It also allows us to evaluate new models as they are
made available. Finally, it also means we can allow others to access our
full workflow. We use LLM-as-a-judge with per-category evaluation.

We use our benchmark to calibrate the auto-grader. This meant dividing
the benchmark into training and test sets. We then had the auto-grader
mark the papers in the benchmark training set and compared the marks
generated by the auto-grader with the marks provided by the humans in
the benchmark. Where there were differences we changed the prompts and
settings of the models that were doing the evaluation. We also made more
substantial changes such as providing each aspect of the submission
separately. The final evaluation was done on all projects, not just the
test set. We graded each project ten times, and the mean standard
deviation was 1.2 points out of 45. This produced an automated grading
pipeline that provides an overall mark within 10 per cent of the
human-provided total mark, and not substantially different on any
particular category. Grading consistency results are in
Appendix~\ref{sec-grading}.

The AI model that underpins our automated grading system is Claude
Sonnet 4.5, with GPT-5.2 used for benchmark validation. Each category is
a separate API call. We prompt the model with the category, the
evaluation criteria, the score options and explicit descriptions, and
the relevant project material. For reproducibility, the model receives
all scripts, the README, and project management files. For writing, it
receives \texttt{paper.qmd}. For figures it receives both the source
code and the rendered PDF as images. The model returns category name,
numeric score, and a brief comment, as semicolon-separated values. The
system prompt specifies the need for publication-level standards, to
avoid being influenced by jargon or verbose writing, and that the model
should return a lower mark if it is not sure between two options.
Table~\ref{tbl-grading-inputs} specifies the inputs provided to the
grading model for each evaluation type.

\begin{table}

\caption{\label{tbl-grading-inputs}Input materials provided to the LLM
grading model for each evaluation type. Each category is evaluated
individually to reduce context and position bias. Reproducibility
evaluation requires the complete codebase; Presentation and Analysis
require only the manuscript source; Graphs evaluation includes rendered
PDF images to assess visual quality.}

\centering{

\centering
\begin{tabular}[t]{lll}
\toprule
Group & Input materials & File types\\
\midrule
Reproducibility & Scripts, README, requirements.txt & .py, .txt, .md\\
Presentation & Paper source & .qmd\\
Analysis & Paper source & .qmd\\
Graphs & Figure source, rendered output & .qmd, .pdf (as images)\\
Referencing & Paper source, bibliography & .qmd, .bib\\
Writing & Paper source & .qmd\\
\bottomrule
\end{tabular}

}

\end{table}%

\subsection{Project generation pipeline}\label{sec-pipeline}

Having established all this infrastructure we can now have the models
generate projects. This involves prompting each model to generate
projects using a nine-step pipeline (Table~\ref{tbl-pipeline}). Each
step is a separate API call. We do this because we often found the
models got stuck on some component that prevented them from being able
to return a complete project. The history maintains context, but this
does mean that we are erring on the side of providing the models with
every opportunity to succeed in the task. A more strict evaluation would
remove this separation. The models receive the relevant portion of the
rubric for each step. The full rubric is in Appendix~\ref{sec-rubric}.

\begin{table}

\caption{\label{tbl-pipeline}Nine-step pipeline for LLM-generated data
analysis projects. Each step is a separate API call with conversation
history maintained across steps. The pipeline progresses from dataset
selection through code generation, quality assurance, and final
documentation.}

\centering{

\centering
\begin{tabular}[t]{lll}
\toprule
Phase & Step & Output\\
\midrule
 & 1. Planning & Dataset selection, research question\\
\cmidrule{2-3}
 & 2. Download & 01-download.py\\
\cmidrule{2-3}
\multirow[t]{-3}{*}{\raggedright\arraybackslash Data acquisition} & 3. Clean & 02-clean.py\\
\cmidrule{1-3}
 & 4. Testing & 03-test.py (pytest suite)\\
\cmidrule{2-3}
\multirow[t]{-2}{*}{\raggedright\arraybackslash Quality assurance} & 5. Requirements & requirements.txt or uv files\\
\cmidrule{1-3}
 & 6. References & references.bib\\
\cmidrule{2-3}
 & 7. Figures & figures.qmd (visualizations)\\
\cmidrule{2-3}
 & 8. Paper & paper.qmd (final manuscript)\\
\cmidrule{2-3}
\multirow[t]{-4}{*}{\raggedright\arraybackslash Documentation} & 9. README & README.md (reproduction guide)\\
\bottomrule
\end{tabular}

}

\end{table}%

The first phase is data acquisition and comprises three steps: planning,
data download, and data cleaning. In Step 1 (planning), we provide the
model with three datasets from Open Data Toronto from which it can pick:
light duty city vehicle utilization data, deaths of people experiencing
homelessness, and neighbourhood profiles. This is more specific than
what is provided to the students, because we found that the models often
struggled to pick an appropriate dataset and this resulted in them being
unable to do the rest of the process. We instruct the model to select
one, develop a research question, and outline an analytical approach. In
Step 2 (data download), we ask the model to write a Python script to
download the selected dataset. This requires more than just pasting the
dataset URL into a function because sometimes there are multiple
datasets and the model must identify which one to use. The script saves
data to \texttt{data/raw/}. In Step 3 (data cleaning), the model writes
a cleaning script. To inform variable handling, the model receives the
first two lines of each downloaded file. The script saves the cleaned
and prepared dataset to \texttt{data/clean/}.

The second phase is quality assurance and reproducibility and involves
two steps. In Step 4 (testing), the model creates a test suite, often
using \texttt{pydantic} to verify data integrity, check columns, and
data types, and validate value ranges. In Step 5 (requirements), the
model is to deal with package management by using uv and generating a UV
lock file and \texttt{pyproject.toml}.

The third phase involves four steps. In Step 6 (references), the model
creates a \texttt{BibTeX} file with the papers that were cited, as well
as citing the data source and software such as Python packages, as
appropriate. In Step 7 (figures), the model creates a Quarto document
(\texttt{figures.qmd}) generating visualizations. The pipeline renders
this, and the model receives both source and output PDF. In Step 8
(paper), the model writes the final paper as a Quarto document,
receiving the BibTeX file and rendered figures to include as needed. The
paper must include title, abstract, introduction, data description,
results, discussion, and references. In Step 9 (README), the model
creates documentation explaining the research question, repository
structure, and reproduction instructions. Generation costs are
summarized in Appendix~\ref{sec-costs}.

\newpage

\appendix

\section*{Appendix}\label{appendix}
\addcontentsline{toc}{section}{Appendix}

\section{Rubric details}\label{sec-rubric}

The rubric has 17 categories grouped into six evaluation types, each
with explicit criteria per score level (Table~\ref{tbl-full-rubric}).
For example, Measurement (2 points) awards 2 for ``thorough and
specific'' explanation of how phenomena become data entries, 1 for
``some explanation but lacks depth,'' and 0 for ``little or no
explanation.''

\begingroup\fontsize{8}{10}\selectfont

\begin{longtable}[t]{>{\raggedright\arraybackslash}p{1.5cm}>{\raggedright\arraybackslash}p{2cm}>{\raggedleft\arraybackslash}p{0.8cm}>{\raggedright\arraybackslash}p{5cm}>{\raggedright\arraybackslash}p{5cm}}

\caption{\label{tbl-full-rubric}Complete grading rubric for end-to-end
data analysis projects. The rubric has 17 categories grouped into six
evaluation types (Reproducibility, Presentation, Analysis, Graphs,
Referencing, Writing), totaling 45 points. Each category includes
specific evaluation criteria and scoring descriptions for each point
level.}

\tabularnewline

\toprule
Type & Category & Points & Key Criteria & Scoring\\
\midrule
\endfirsthead
\multicolumn{5}{@{}l}{\textit{(continued)}}\\
\toprule
Type & Category & Points & Key Criteria & Scoring\\
\midrule
\endhead

\endfoot
\bottomrule
\endlastfoot
 & Reproducible Workflow & 5 & Code is thoroughly documented code, including preamble, comments, nice structure, and is styled. Hard-coded paths are avoided. A detailed README.md... & 0 – ‘Disorganized; lacks comments; not reproducible’ 1 – ‘Partial structure; some documentation; hard to reproduce’ 2 – ‘Missing two or more major components; still somewhat reproducible’ 3 – 'Missing one major component; still somewhat reproducible' 4 – ‘Clear structure and documentation; readable and reproducible with minor friction’ 5 – ‘Fully reproducible; all components present'\\
\cmidrule{2-5}\nopagebreak
 & Tests & 3 & Test suite exists in its own script Tests go beyond basic null/invalid checks Tests use proper framework (unittest or pytest) Test suite is extensi... & 0 – ‘No or few tests’ 1 – ‘Basic or minimal tests written’ 2 – ‘Organized suite covering some major components with structure’ 3 – ‘Extensive and sophisticated test suite using frameworks like pytest or unittest’\\
\cmidrule{2-5}\nopagebreak
\multirow[t]{-3}{1.5cm}{\raggedright\arraybackslash Reproducibility} & Author, Date, \& Repo & 1 & Author name is clearly included Date of submission in unambiguous format is included Link to GitHub repo is included & 0 – ‘One or more elements missing or unclear’ 1 – ‘All three (author, date, GitHub repo) are clearly included’\\
\cmidrule{1-5}\pagebreak[0]
 & Title & 2 & Title is informative (not generic like "Paper X") Title explains the story Subtitle conveys the main finding No puns used & 0 – ‘Missing or generic’ 1 – ‘Informative but not fully representative of paper’ 2 – ‘Clear, specific, and polished. Includes subtitle with key finding’\\
\cmidrule{2-5}\nopagebreak
 & Abstract & 4 & Pitched to non-specialist audience Explains what was done Explains what was found Explains why this matters Includes scene setting States the resea... & 0 – ‘Missing or lacks two or more required elements’ 1 – ‘Most elements present but unclear or inaccessible to general audience’ 2 – ‘Clear and complete but too wordy, or technical’ 3 – ‘Clear, concise (\textasciitilde{}4 sentences), pitched to general audience’ 4 – ‘All elements covered in clear, concise prose with relevance highlighted’\\
\cmidrule{2-5}\nopagebreak
\multirow[t]{-3}{1.5cm}{\raggedright\arraybackslash Presentation} & Introduction & 4 & Provides broader context to motivate Details what the paper is about Identifies a clear gap that needs to be filled States what was done States wha... & 0 – ‘Missing or incomplete’ 1 – ‘Some elements included, but lacks clarity or structure’ 2 – ‘Most elements present, some issues with clarity’ 3 – ‘Clear and logically structured’ 4 – ‘Fully self-contained; explains motivation, findings, and structure’\\
\cmidrule{1-5}\pagebreak[0]
 & Measurement & 2 & Explains how real-world phenomena becomes data entry Describes collection method Describes measurement method Describes categorization/coding metho... & 0 – ‘Vague or missing’ 1 – ‘Some explanation’ 2 – ‘Thorough and specific’\\
\cmidrule{2-5}\nopagebreak
 & Data & 4 & Discusses alternative datasets and why not used Mentions if variables were constructed Notes high-level cleaning aspects Focuses on destination, no... & 0 – ‘Missing or incoherent’ 1 – ‘Described but with major omissions’ 2 – ‘Adequate with minor clarity or rationale issues’ 3 – ‘All variables of interest explained; some polish missing’ 4 – ‘Thorough and clearly justified; All variables of interest are clearly explained’\\
\cmidrule{2-5}\nopagebreak
\multirow[t]{-3}{1.5cm}{\raggedright\arraybackslash Analysis} & Discussion & 4 & 2-3 sections (one per key takeaway) Each takeaway clearly highlighted \& explained in detail Explains what results mean relative to research questio... & 0 – ‘Results restated without meaning’ 1 - “Unclear interpretation’ 2 – ‘Interpretation present but shallow’ 3 – ‘Results explained in context with some weak points’ 4 – ‘Strong, thoughtful interpretation linked to findings and limitations’\\
\cmidrule{1-5}\pagebreak[0]
 & Data Tables & 2 & Graphs and/or tables showing what data looks like Summary statistics table is included Table contains data.describe() or equivalent Located in the ... & 0 – ‘Missing or insufficient’ 1 – ‘Table is present in the DATA SECTION, but it does not contain summary statistics.’ 2 – ‘There is a table in the DATA SECTION containing summary statistics.'\\
\cmidrule{2-5}\nopagebreak
 & Results Figures & 4 & Analytically focused graphs/tables present Journal-quality appearance Clear and digestible Serves clear purpose and supports claims Fully self-cont... & 0 – ‘Missing or unreadable’ 1 – ‘One or more present but poorly formatted or unclear’ 2 – ‘Clear and interpretable’ 3 – ‘Well-designed and polished’ 4 – ‘Clear, polished, labeled, and directly supports claims’\\
\cmidrule{2-5}\nopagebreak
\multirow[t]{-3}{1.5cm}{\raggedright\arraybackslash Graphs} & Captions & 2 & All figures and tables have captions Captions are detailed and meaningful Main point clear without reading text Does NOT simply state figure type (... & 0 – ‘Most missing or meaningless’ 1 – ‘Present but not sufficiently detailed’ 2 – ‘All figures/tables have meaningful, standalone captions’\\
\cmidrule{1-5}\pagebreak[0]
 & Cross References & 1 & All figures and tables are labeled All figures and tables referenced in text (e.g., @fig-x) & 0 – ‘No cross-references or missing cross-references’ 1 – ‘All visuals and equations numbered and referred to in the text’\\
\cmidrule{2-5}\nopagebreak
 & Referencing & 2 & All citations correctly capitalized and formatted No reversed names (e.g., "Data, Toronto Open" should be "Open Data Toronto") & 0 – ‘Poor or missing citations’ 1 – ‘Citations have minor formatting issues’ 2 – ‘Complete and correctly formatted citations’\\
\cmidrule{2-5}\nopagebreak
\multirow[t]{-3}{1.5cm}{\raggedright\arraybackslash Referencing} & Python Cited & 1 & Python given in text citation Python appears in reference list Python packages cited appropriately & 0 – ‘Python is not cited’ 1 – ‘Python is cited’\\
\cmidrule{1-5}\pagebreak[0]
 & Professionality & 3 & Free of noticeable typos Grammatically correct Coherent, concise and clear Mature tone, not overly technical or wordy & 0 – ‘Numerous grammatical/spelling issues’ 1 – ‘Readable but with issues in clarity or flow’ 2 – ‘Clear and mostly error-free’ 3 – ‘Polished and mature’\\
\cmidrule{2-5}\nopagebreak
\multirow[t]{-2}{1.5cm}{\raggedright\arraybackslash Writing} & Banned Words & 1 & check if the text contains any of the following words: advanced, comprehensive, critical, crucial, data-driven, delve/s, imperative, insight/s & 0 - ‘One or more of the words are used on the list’ 1- ‘None of the words on the list are used’\\*

\end{longtable}

\endgroup{}

\newpage

\section{Per-project scores}\label{sec-per-project}

\begin{figure}

\centering{

\pandocbounded{\includegraphics[keepaspectratio]{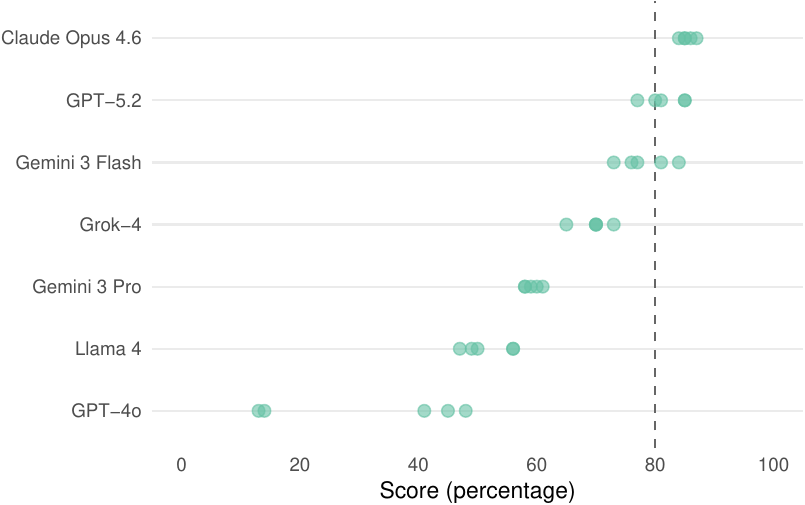}}

}

\caption{\label{fig-per-project}Distribution of individual project
scores for each model. Each point represents one of five projects.
Models are ordered by mean score.}

\end{figure}

The main results are average scores across five projects per model.
Figure~\ref{fig-per-project} shows the individual project scores. Claude
Opus 4.6 was the most consistent, with all five projects scoring between
84 and 87 per cent. GPT-4o had the highest variance, ranging from 13 to
48 per cent.

\newpage

\section{Example generated papers}\label{sec-examples}

\begin{figure}

\begin{minipage}{0.33\linewidth}

\centering{

\includegraphics[width=1\linewidth,height=\textheight,keepaspectratio]{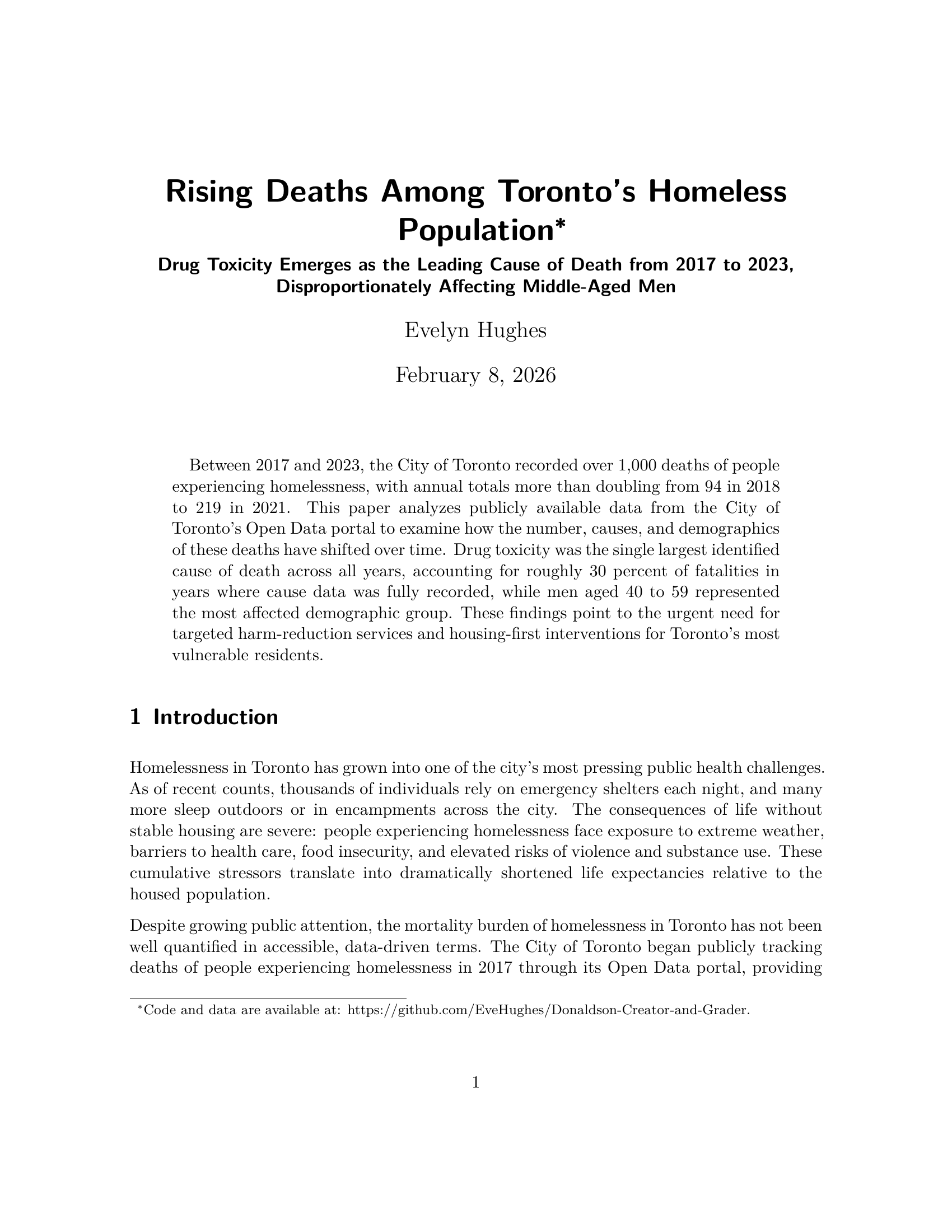}

}

\subcaption{\label{fig-ex-opus}Claude Opus 4.6, Project 5 (87\%)}

\end{minipage}%
\begin{minipage}{0.33\linewidth}

\centering{

\includegraphics[width=1\linewidth,height=\textheight,keepaspectratio]{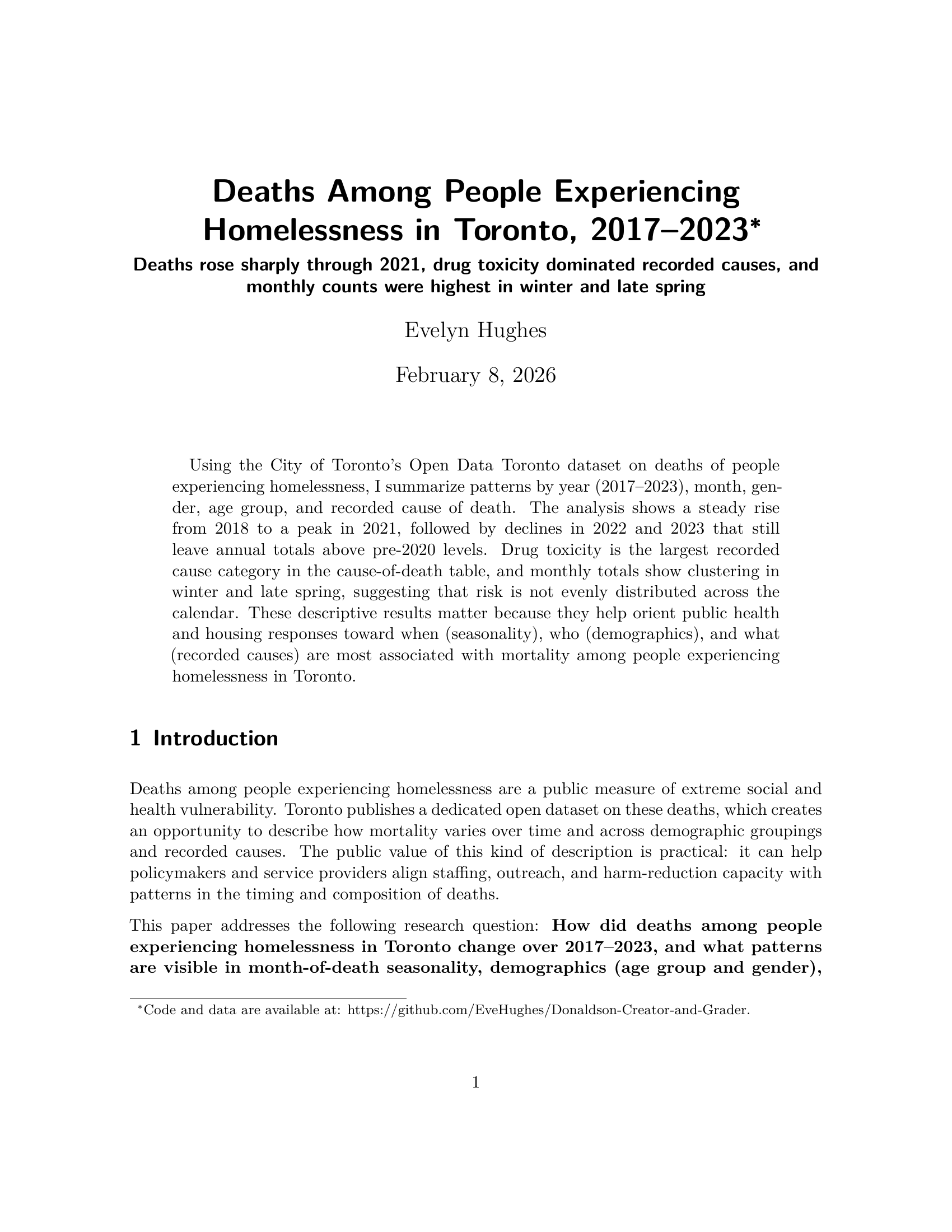}

}

\subcaption{\label{fig-ex-gpt52}GPT-5.2, Project 2 (85\%)}

\end{minipage}%
\begin{minipage}{0.33\linewidth}

\centering{

\includegraphics[width=1\linewidth,height=\textheight,keepaspectratio]{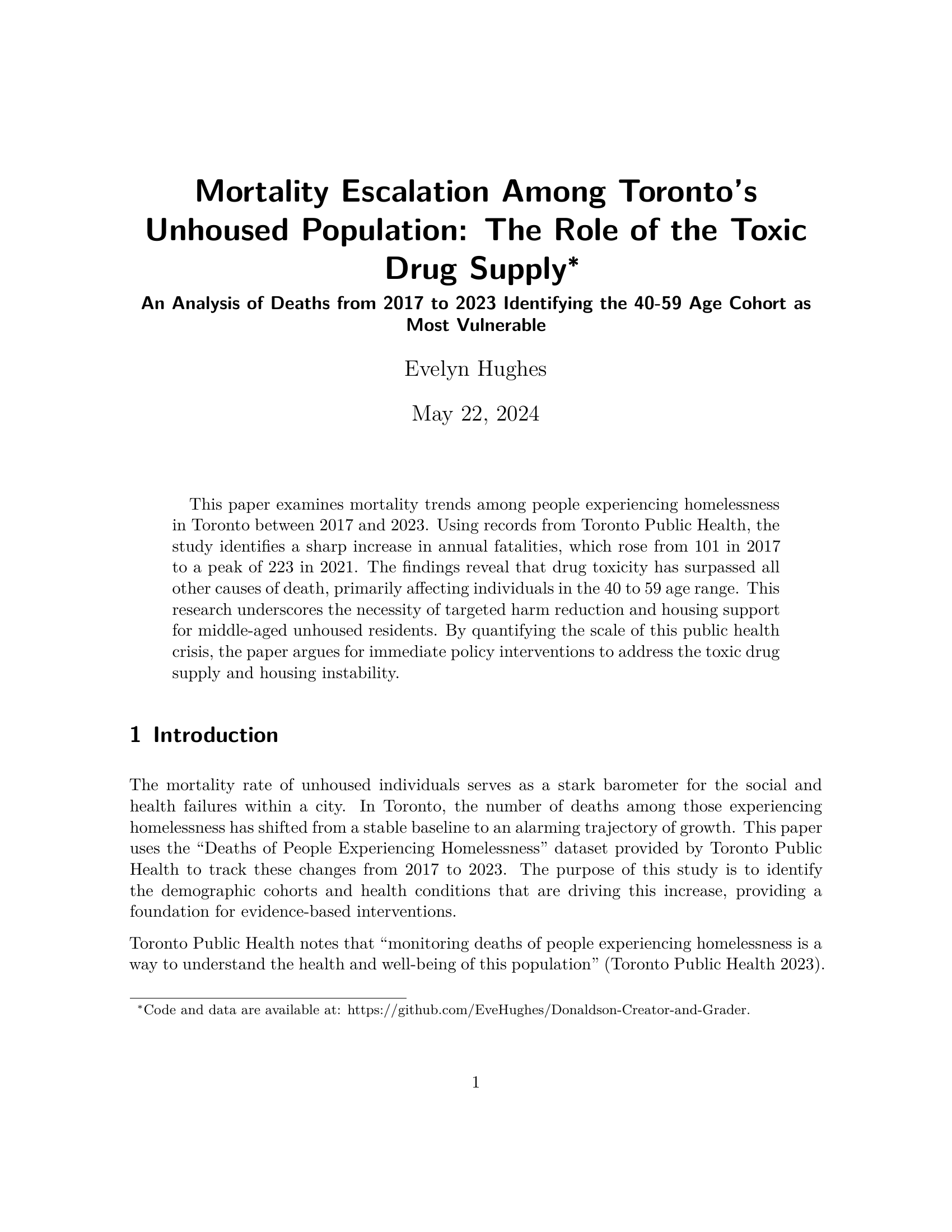}

}

\subcaption{\label{fig-ex-gemini}Gemini 3 Flash, Project 3 (84\%)}

\end{minipage}%
\newline
\begin{minipage}{0.33\linewidth}

\centering{

\includegraphics[width=1\linewidth,height=\textheight,keepaspectratio]{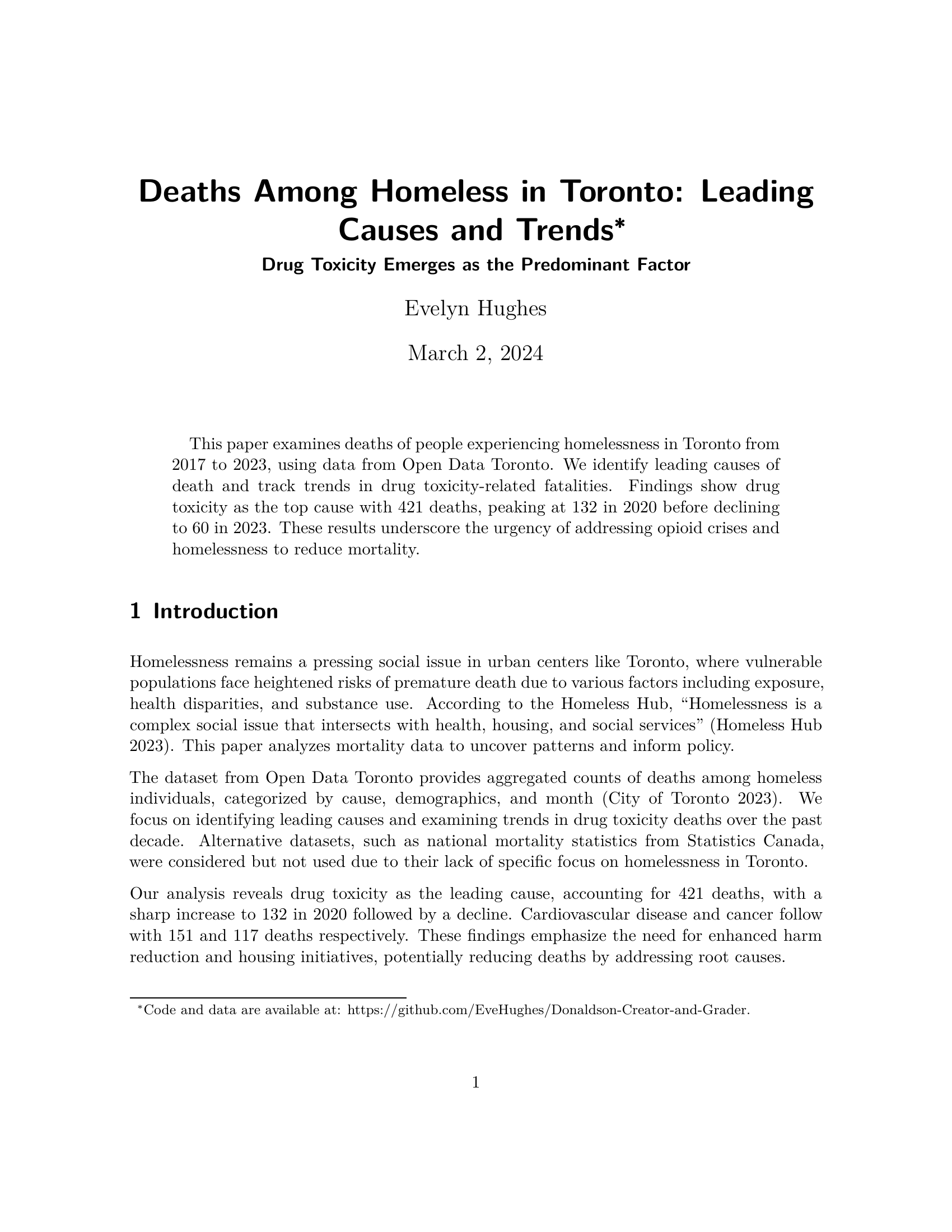}

}

\subcaption{\label{fig-ex-grok}Grok-4, Project 5 (73\%)}

\end{minipage}%
\begin{minipage}{0.33\linewidth}

\centering{

\includegraphics[width=1\linewidth,height=\textheight,keepaspectratio]{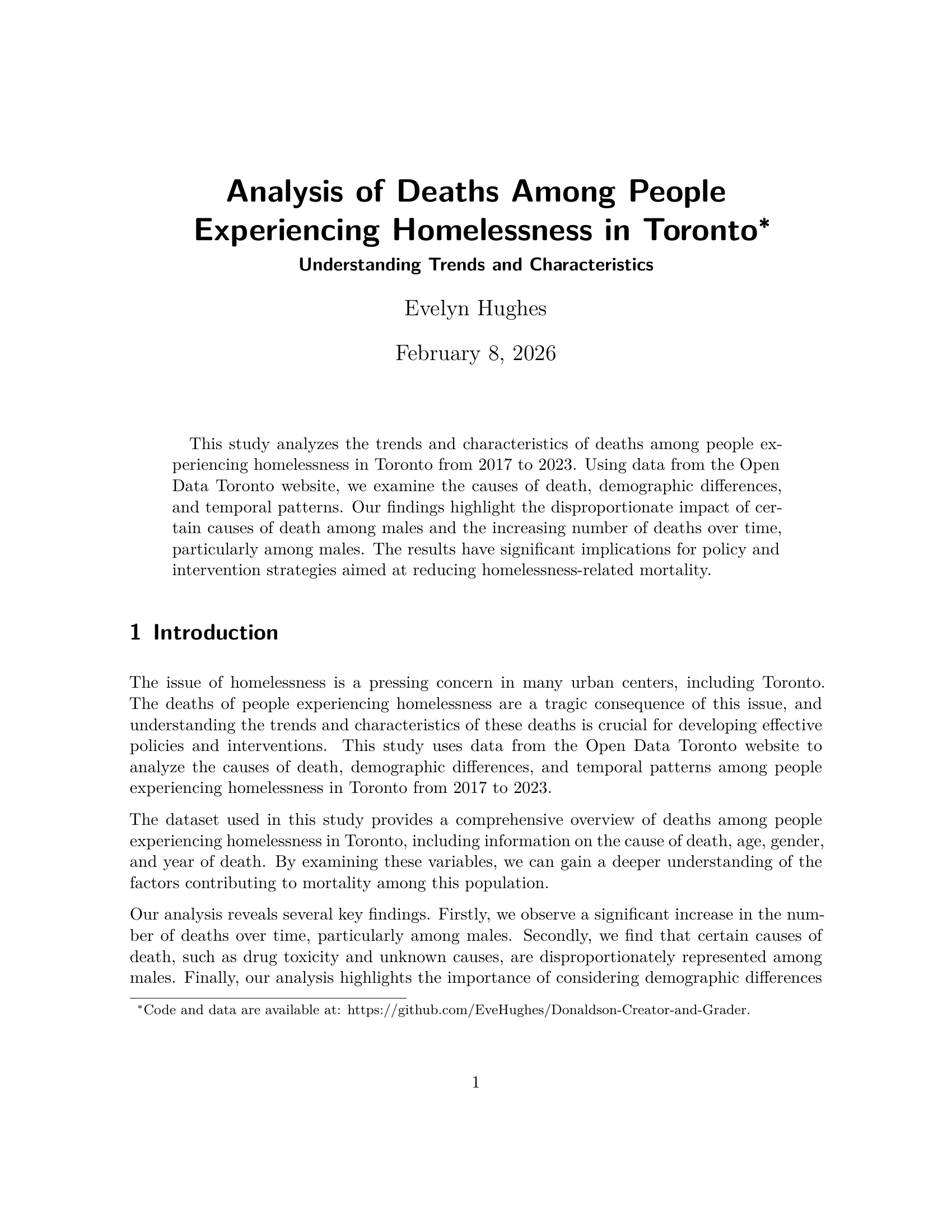}

}

\subcaption{\label{fig-ex-llama}Llama 4, Project 5 (47\%)}

\end{minipage}%
\begin{minipage}{0.33\linewidth}

\centering{

\includegraphics[width=1\linewidth,height=\textheight,keepaspectratio]{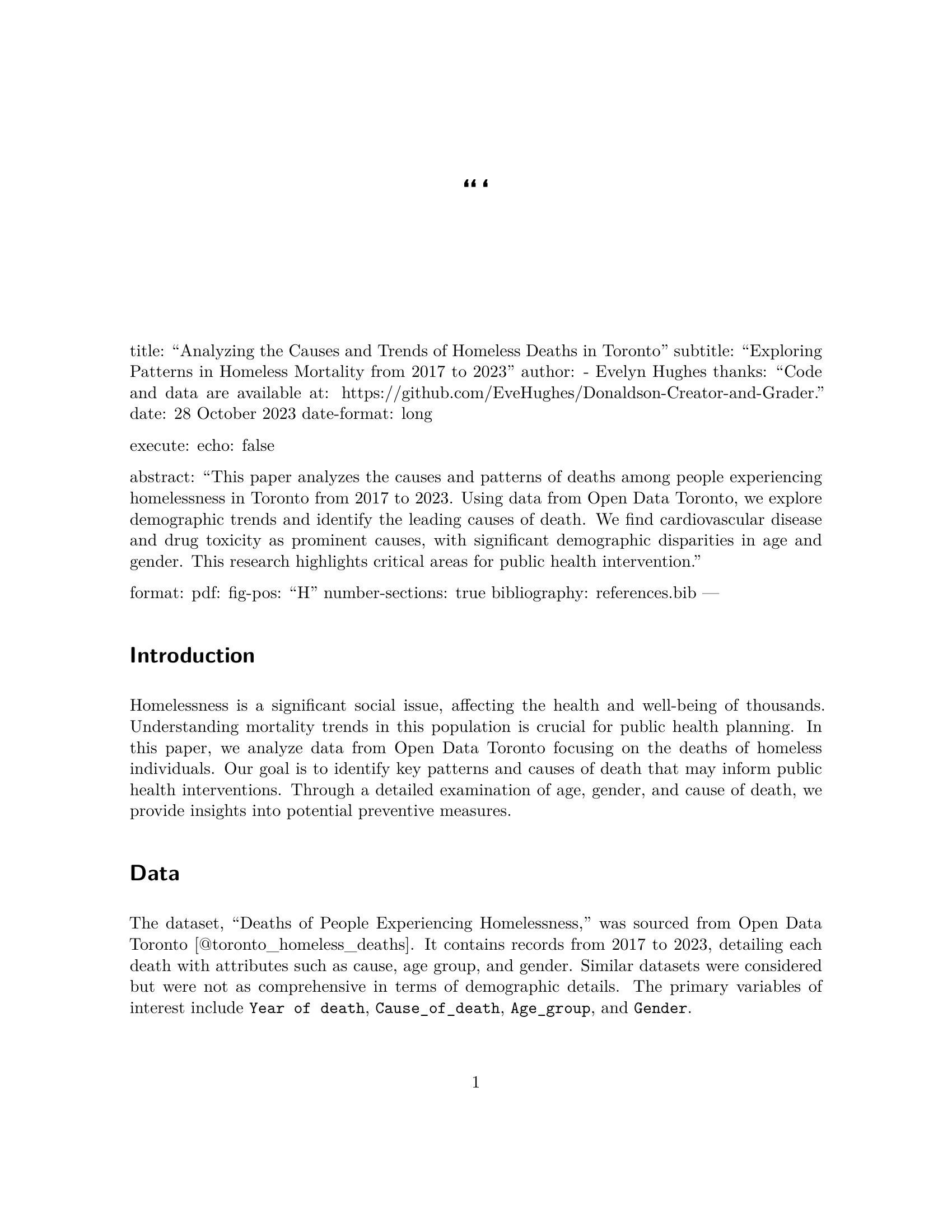}

}

\subcaption{\label{fig-ex-gpt4o}GPT-4o, Project 2 (13\%)}

\end{minipage}%

\caption{\label{fig-examples-top}First pages of six generated papers.}

\end{figure}

To illustrate the range of output quality Figure~\ref{fig-examples-top}
shows the first page of six generated papers. Higher-scoring papers have
well-formatted titles, abstracts, and introductions. Lower-scoring
papers have generic writing (Llama 4) and broken Quarto rendering
(GPT-4o).

\newpage

\section{Grading consistency and validation}\label{sec-grading}

Claude Sonnet 4.5 graded each of the 40 model-generated projects three
times. Table~\ref{tbl-grading-consistency} shows the standard deviation
for each grading run, for each model's five projects, as well as the
average. The overall average standard deviation was 0.8 points out of
45. We include the grading of projects created by Claude Sonnet 4.5 here
for information, but they are not in our main results because we use
Claude Sonnet 4.5 as the grader.

To validate the auto-grader, both Claude Sonnet 4.5 and GPT-5.2
independently graded all 40 human-evaluated benchmark projects three
times each (Table~\ref{tbl-grader-comparison}). The two graders produced
correlated rankings (r = 0.85) but differed in calibration: Claude
Sonnet 4.5 averaged 25.4 out of 45 while GPT-5.2 averaged 20.3. Both
graders were internally consistent, with mean within-project standard
deviations below 1.0 point.

\begin{table}

\caption{\label{tbl-grading-consistency}Grading consistency across three
evaluation runs per project. Each cell shows the standard deviation of
total scores across three independent grading runs for that project. Low
values indicate the auto-grader assigns similar scores when
re-evaluating the same project.}

\centering{

\centering
\resizebox{\ifdim\width>\linewidth\linewidth\else\width\fi}{!}{
\begin{tabular}[t]{lrrrrrr}
\toprule
Model & Project 1 & Project 2 & Project 3 & Project 4 & Project 5 & Mean\\
\midrule
Claude Opus 4.6 & 0.6 & 1.0 & 1.5 & 1.5 & 0.0 & 0.9\\
Claude Sonnet 4.5 & 0.6 & 1.2 & 1.5 & 2.1 & 0.6 & 1.2\\
Gemini 3 Flash & 0.0 & 0.6 & 1.2 & 0.0 & 0.6 & 0.5\\
Gemini 3 Pro & 0.0 & 0.0 & 0.6 & 1.5 & 0.0 & 0.4\\
GPT-4o & 1.2 & 0.6 & 1.2 & 0.0 & 0.6 & 0.7\\
GPT-5.2 & 0.6 & 0.6 & 1.7 & 1.2 & 0.6 & 0.9\\
Grok-4 & 2.1 & 1.0 & 1.5 & 0.6 & 0.6 & 1.2\\
Llama 4 & 1.2 & 1.0 & 1.0 & 0.0 & 0.0 & 0.6\\
\bottomrule
\end{tabular}}

}

\end{table}%

\begin{table}

\caption{\label{tbl-grader-comparison}Comparison of two auto-graders on
the 40 human-evaluated benchmark projects. Both graders evaluated each
project three times. The graders show high rank-order agreement (r =
0.85) but differ in overall calibration, with Claude Sonnet 4.5
assigning higher scores on average.}

\centering{

\centering
\begin{tabular}[t]{lrr}
\toprule
Grader & Mean score & SD (within-project)\\
\midrule
Claude Sonnet 4.5 & 25.4 & 0.9\\
GPT-5.2 & 20.3 & 0.7\\
Correlation (r) & 0.85 & —\\
\bottomrule
\end{tabular}

}

\end{table}%

\newpage

\section{Generation costs}\label{sec-costs}

Table~\ref{tbl-creation-costs} summarizes the computational cost of
generating the five projects for each model. Total cost includes all API
calls across the nine-step pipeline. Retry attempts occur when a step
fails (e.g.~a script produces an error) and the pipeline re-prompts the
model with the traceback. Llama 4 and Gemini 3 Pro required the most
retries (24 and 22), which is consistent with their lower scores.
Grading all model-generated projects three times each cost \$58.87 using
Claude Sonnet 4.5 (taking 3.9 hours) and \$36.05 using GPT-5.2 (taking
2.7 hours).

\begin{table}

\caption{\label{tbl-creation-costs}Cost and time to generate five
projects per model. Total cost reflects all API calls across the
nine-step generation pipeline. Retry attempts indicate how many pipeline
steps required re-prompting after failure. Models requiring more retries
tended to score lower.}

\centering{

\centering
\begin{tabular}[t]{lrrr}
\toprule
Model & Time (min) & Cost (USD) & Retries\\
\midrule
Claude Opus 4.6 & 42 & \$4.75 & 5\\
Gemini 3 Pro & 73 & \$3.31 & 22\\
Grok-4 & 47 & \$1.73 & 8\\
GPT-5.2 & 34 & \$1.73 & 5\\
GPT-4o & 17 & \$1.06 & 6\\
Gemini 3 Flash & 26 & \$0.39 & 7\\
Llama 4 & 79 & \$0.32 & 24\\
\bottomrule
\end{tabular}

}

\end{table}%

\newpage

\section*{References}\label{references}
\addcontentsline{toc}{section}{References}

\phantomsection\label{refs}


\begin{CSLReferences}{1}{0}
\bibitem[\citeproctext]{ref-alexander2023}
Alexander, Rohan. 2023. \emph{{Telling Stories with Data}}. Chapman;
Hall/CRC. \url{https://tellingstorieswithdata.com}.

\bibitem[\citeproctext]{ref-austin2021}
Austin, Jacob, Augustus Odena, Maxwell Nye, Maarten Bosma, Henryk
Michalewski, David Dohan, Ellen Jiang, et al. 2021. {``Program Synthesis
with Large Language Models.''}
\url{https://doi.org/10.48550/arXiv.2108.07732}.

\bibitem[\citeproctext]{ref-bowman2021}
Bowman, Samuel R., and George Dahl. 2021. {``What Will It Take to Fix
Benchmarking in Natural Language Understanding?''} In \emph{Proceedings
of the 2021 Conference of the North American Chapter of the Association
for Computational Linguistics: Human Language Technologies}, edited by
Kristina Toutanova, Anna Rumshisky, Luke Zettlemoyer, Dilek Hakkani-Tur,
Iz Beltagy, Steven Bethard, Ryan Cotterell, Tanmoy Chakraborty, and
Yichao Zhou, 4843--55. Online: Association for Computational
Linguistics. \url{https://doi.org/10.18653/v1/2021.naacl-main.385}.

\bibitem[\citeproctext]{ref-chen2021}
Chen, Mark, Jerry Tworek, Heewoo Jun, Qiming Yuan, Henrique Ponde de
Oliveira Pinto, Jared Kaplan, Harri Edwards, et al. 2021. {``{Evaluating
Large Language Models Trained on Code}.''}
\url{https://doi.org/10.48550/arXiv.2107.03374}.

\bibitem[\citeproctext]{ref-chen2025}
Chen, Qiang, Tianyang Han, Jin Li, Ye Luo, Yuxiao Wu, Xiaowei Zhang, and
Tuo Zhou. 2025. {``Can AI Master Econometrics? Evidence from
Econometrics AI Agent on Expert-Level Tasks.''}
\url{https://doi.org/10.48550/arXiv.2506.00856}.

\bibitem[\citeproctext]{ref-evkaya2026}
Evkaya, Ozan, and Miguel de Carvalho. 2026. {``Using {ChatGPT} for Data
Science Analyses.''} \emph{Harvard Data Science Review} 8 (1).
\url{https://doi.org/10.1162/99608f92.c9429f07}.

\bibitem[\citeproctext]{ref-gu2025}
Gu, Yang, Hengyu You, Jian Cao, Muran Yu, Haoran Fan, and Shiyou Qian.
2025. {``Large Language Models for Constructing and Optimizing Machine
Learning Workflows: A Survey.''} \emph{ACM Transactions on Software
Engineering and Methodology}, October.
\url{https://doi.org/10.1145/3773084}.

\bibitem[\citeproctext]{ref-jimenez2024}
Jimenez, Carlos E, John Yang, Alexander Wettig, Shunyu Yao, Kexin Pei,
Ofir Press, and Karthik R Narasimhan. 2024. {``{{SWE}-bench: Can
Language Models Resolve Real-world Github Issues?}''} In
\emph{International Conference on Learning Representations}.
\url{https://openreview.net/forum?id=VTF8yNQM66}.

\bibitem[\citeproctext]{ref-Kobak2025}
Kobak, Dmitry, Rita González-Márquez, Emőke-Ágnes Horvát, and Jan Lause.
2025. {``Delving into LLM-Assisted Writing in Biomedical Publications
Through Excess Vocabulary.''} \emph{Science Advances} 11 (27).
\url{https://doi.org/10.1126/sciadv.adt3813}.

\bibitem[\citeproctext]{ref-Kusumegi2025}
Kusumegi, Keigo, Xinyu Yang, Paul Ginsparg, Mathijs de Vaan, Toby
Stuart, and Yian Yin. 2025. {``Scientific Production in the Era of Large
Language Models.''} \emph{Science} 390 (6779): 1240--43.
\url{https://doi.org/10.1126/science.adw3000}.

\bibitem[\citeproctext]{ref-lai2023}
Lai, Yuhang, Chengxi Li, Yiming Wang, Tianyi Zhang, Ruiqi Zhong, Luke
Zettlemoyer, Wen-tau Yih, Daniel Fried, Sida Wang, and Tao Yu. 2023.
{``DS-1000: A Natural and Reliable Benchmark for Data Science Code
Generation.''} In \emph{Proceedings of the 40th International Conference
on Machine Learning}. ICML'23. Honolulu, Hawaii, USA: JMLR.org.
\url{https://dl.acm.org/doi/10.5555/3618408.3619164}.

\bibitem[\citeproctext]{ref-nasem2018}
National Academies of Sciences, Engineering, and Medicine. 2018.
{``{Data Science for Undergraduates: Opportunities and Options}.''}
\url{https://doi.org/10.17226/25104}.

\bibitem[\citeproctext]{ref-Noy2023}
Noy, Shakked, and Whitney Zhang. 2023. {``Experimental Evidence on the
Productivity Effects of Generative Artificial Intelligence.''}
\emph{Science} 381 (6654): 187--92.
\url{https://doi.org/10.1126/science.adh2586}.

\bibitem[\citeproctext]{ref-citeR}
R Core Team. 2025. \emph{R: A Language and Environment for Statistical
Computing}. Vienna, Austria: R Foundation for Statistical Computing.
\url{https://www.R-project.org/}.

\bibitem[\citeproctext]{ref-raji2021}
Raji, Deborah, Emily Denton, Emily M. Bender, Alex Hanna, and
Amandalynne Paullada. 2021. {``AI and the Everything in the Whole Wide
World Benchmark.''} In \emph{Proceedings of the Neural Information
Processing Systems Track on Datasets and Benchmarks}, edited by J.
Vanschoren and S. Yeung.
\url{https://doi.org/10.48550/arXiv.2111.15366}.

\bibitem[\citeproctext]{ref-ranganathan2026}
Ranganathan, Aruna, and Xingqi Maggie Ye. 2026. {``{AI} Doesn't Reduce
Work---It Intensifies It.''} \emph{Harvard Business Review}, February.
\url{https://hbr.org/2026/02/ai-doesnt-reduce-work-it-intensifies-it}.

\bibitem[\citeproctext]{ref-reuel2024}
Reuel, Anka, Amelia Hardy, Chandler Smith, Max Lamparth, Malcolm Hardy,
and Mykel J. Kochenderfer. 2024. {``BetterBench: Assessing AI
Benchmarks, Uncovering Issues, and Establishing Best Practices.''} In
\emph{Proceedings of the 38th International Conference on Neural
Information Processing Systems}. NIPS '24. Vancouver, BC, Canada.
\url{https://doi.org/10.48550/arXiv.2411.12990}.

\bibitem[\citeproctext]{ref-robinsonnolis2020}
Robinson, Emily, and Jacqueline Nolis. 2020. \emph{Build a Career in
Data Science}. Manning Publications.
\url{https://www.manning.com/books/build-a-career-in-data-science}.

\bibitem[\citeproctext]{ref-singh2025}
Singh, Shivalika, Yiyang Nan, Alex Wang, Daniel D'Souza, Sayash Kapoor,
Ahmet Üstün, Sanmi Koyejo, et al. 2025. {``The Leaderboard Illusion.''}
\url{https://doi.org/10.48550/arXiv.2504.20879}.

\bibitem[\citeproctext]{ref-szymanski2024}
Szymanski, Annalisa, Noah Ziems, Heather A. Eicher-Miller, Toby Jia-Jun
Li, Meng Jiang, and Ronald A. Metoyer. 2024. {``{Limitations of the
LLM-as-a-Judge Approach for Evaluating LLM Outputs in Expert Knowledge
Tasks}.''} \url{https://doi.org/10.48550/arXiv.2410.20266}.

\bibitem[\citeproctext]{ref-timbers2022}
Timbers, Tiffany, Trevor Campbell, and Melissa Lee. 2022. \emph{{Data
Science: A First Introduction}}. Chapman; Hall/CRC.
\url{https://datasciencebook.ca}.

\bibitem[\citeproctext]{ref-tidyverse}
Wickham, Hadley, Mara Averick, Jennifer Bryan, Winston Chang, Lucy
D'Agostino McGowan, Romain François, Garrett Grolemund, et al. 2019.
{``Welcome to the {Tidyverse}.''} \emph{Journal of Open Source Software}
4 (43): 1686. \url{https://doi.org/10.21105/joss.01686}.

\bibitem[\citeproctext]{ref-zhang2025}
Zhang, Dan, Sining Zhoubian, Min Cai, Fengzu Li, Lekang Yang, Wei Wang,
Tianjiao Dong, Ziniu Hu, Jie Tang, and Yisong Yue. 2025.
{``DataSciBench: An LLM Agent Benchmark for Data Science.''}
\url{https://doi.org/10.48550/arXiv.2502.13897}.

\bibitem[\citeproctext]{ref-zheng2023}
Zheng, Lianmin, Wei-Lin Chiang, Ying Sheng, Siyuan Zhuang, Zhanghao Wu,
Yonghao Zhuang, Zi Lin, et al. 2023. {``{Judging LLM-as-a-Judge with
MT-Bench and Chatbot Arena}.''} In \emph{Advances in Neural Information
Processing Systems}, edited by A. Oh, T. Naumann, A. Globerson, K.
Saenko, M. Hardt, and S. Levine, 36:46595--623.
\url{https://doi.org/10.48550/arXiv.2306.05685}.

\bibitem[\citeproctext]{ref-zhu2025}
Zhu, Lianghui, Xinggang Wang, and Xinlong Wang. 2025. {``Judge{LM}:
Fine-Tuned Large Language Models Are Scalable Judges.''} In \emph{The
Thirteenth International Conference on Learning Representations}.
\url{https://openreview.net/forum?id=xsELpEPn4A}.

\bibitem[\citeproctext]{ref-zhuo2025}
Zhuo, Terry Yue, Vu Minh Chien, Jenny Chim, Han Hu, Wenhao Yu, Ratnadira
Widyasari, Imam Nur Bani Yusuf, et al. 2025. {``BigCodeBench:
Benchmarking Code Generation with Diverse Function Calls and Complex
Instructions.''} In \emph{The Thirteenth International Conference on
Learning Representations}.
\url{https://openreview.net/forum?id=YrycTjllL0}.

\end{CSLReferences}
\end{document}